\documentclass[epsf,twocolumn,showpacs,preprintnumbers]{revtex4}
\usepackage{graphics}
\usepackage{graphicx}
\usepackage{dcolumn} 
\usepackage{bm}
\usepackage{color}
\usepackage{epsfig}
\begin{document}
\title{Spin-orbit interaction driven collective 
 electron-hole excitations \\
 in a noncentrosymmetric nodal loop Weyl semimetal
}
\author{Kyo-Hoon Ahn$^1$}
\author{Kwan-Woo Lee$^{1,2}$}
\email{mckwan@korea.ac.kr} 
\author{Warren E. Pickett$^3$}
\email{pickett@physics.ucdavis.edu}
\affiliation{
 $^1$Department of Applied Physics, Graduate School, Korea University, Sejong 339-700, Korea\\
 $^2$ Department of Display and Semiconductor Physics, Korea University, Sejong 339-700, Korea \\
$^3$Department of Physics, University of California, Davis,
  CA 95616, USA
}
\date{\today}
\pacs{71.20.Be,71.18.+y,78.20.Ci,72.15.Eb} 
\begin{abstract}
NbP is one member of a new class of nodal loop semimetals characterized by the
cooperative effects of spin-orbit coupling (SOC) and a lack of inversion center.
Here transport and spectroscopic properties of NbP are evaluated
using density functional theory methods. 
SOC together with the lack of inversion symmetry splits degeneracies, 
giving rise to ``Russian
doll nested'' Fermi surfaces containing 4$\times$10$^{-4}$ electron (hole)
carriers/f.u. Due to the modest SOC strength in Nb, the Fermi surfaces
map out the Weyl nodal loops. Calculated structure around T$^*$$\approx$100 K in
transport properties
reproduces well the observed transport behavior only when SOC is included, 
attesting to the precision
of the (delicate) calculations and the stoichiometry of the samples.  
Low energy collective electron-hole excitations (plasmons) 
in the 20-60 meV range result from the nodal loop splitting.
\end{abstract}
\maketitle

\begin{figure}[tbp]
{\resizebox{6cm}{6cm}{\includegraphics{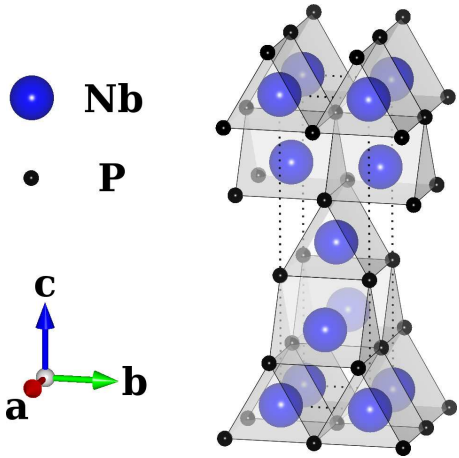}}}
\caption{(Color online) Body-centered tetragonal $I4_1md$ structure of NbP.
It consists of edge-sharing NbP$_6$ trigonal prisms, which are 
90$^\circ$ rotated
in the next layer. The lack of inversion symmetry is evident: the prisms
all ``point'' upward.
}
\label{str}
\end{figure}

\section{Introduction}
Weyl semimetals (WSs) have Weyl points, comprised of
a linear crossing of valence and conduction bands\cite{wan} near the Fermi energy E$_F$. 
WSs can be realized from a lack of time-reversal symmetry 
(magnetic ordering, or applied field $B$),\cite{wan,gxu,burkov}
or from no inversion symmetry in three-dimensional compounds,\cite{weng,huang} and
show peculiar features, such as Fermi arcs from surface states 
connecting two Weyl points with
different chirality.\cite{wan,weng,huang} Experimental evidence has been obtained
using x-ray angle-resolved photoemission spectroscopy.\cite{Lv} 
Extremely large magnetoresistivity (XMR) also occurs in some WSs, 
with a recent analysis by Burkov\cite{burkov} concluding that XMR should be
expected in high mobility semimetals where Weyl or Dirac points lie
near the Fermi surface (FS). 
XMR was observed also in the semimetal WTe$_2$, showing unsaturated MR
up to 60 T,\cite{cava14} attributed to precise electron and hole carrier compensation,
as in Bi and other stoichiometric semimetals.

Recently a ``nodal loop'' Weyl semimetal phase with potentially distinctive properties
has been proposed on theoretical 
grounds\cite{weng,huang} in the nonmagnetic non-centrosymmetric isovalent 
class TaAs, TaP, NbAs, and NbP, whose structure is shown in Fig. 1. 
A previous proposal of nodal loop WSs by 
Phillips and Aji invoked time-reversal symmetry breaking\cite{aji} 
rather than the
lack of inversion symmetry of this class of pnictides. 
Since these suggestions, XMR has been observed in both TaAs\cite{czhang,czhang2} 
and NbP.\cite{she,czhang2}
NbP, upon which we focus here, shows XMR of 8,500 at 2 K
for $B=9$ T, increasing to 
$3\times 10^4$ at $4$ K at $B$=30 T, to 10$^5$ at 60 T,
and remains unsaturated.\cite{she} These colossal XMR values reflect the excellent
conduction at low temperatures, with resistivity as low as 0.1$\mu\Omega$ cm, and
a metal to insulator crossover with field at 100 K and below. 
A high mobility $\mu$=$5\times 10^6$ cm$^2$/V-s 
and low carrier density of $1.5\times 10^{18}$ cm$^{-3}$ at low $T$
was inferred from the
conventional single band expressions, which however are not quantitative 
for multiband semimetals.
As temperature increases, both the Hall coefficient $R_H$ changes sign from electron-like
to hole-like and the mobility 
rapidly changes in the 50-150 K range, correlating with the metal to insulator
crossover. A multiband model and 
additional data will be necessary to sort out
individual carrier densities and mobilities.  
NbP has the smallest spin-orbit coupling (SOC) of this class, making it the most
delicate example of the features of nodal loop WSs. 

From both theoretical and experimental viewpoints, nodal loop semimetals comprise a new
class of materials with topological behavior.
In this paper, we focus on effects of SOC on their electronic
properties and evaluate the
optical and transport coefficients of NbP, using density functional theory based methods
that enable a direct theory-experiment comparison.
A central feature that emerges is electronic fine structure driven by the combination
of SOC and lack of an inversion center, a feature elaborated on by 
Samokhin\cite{samokhin} from a model viewpoint.
This splitting in a semimetal with small closed Fermi surfaces
(FSs) results in a ``Russian doll nested'' pair of
FSs, and new low energy interband transition strength pilfered from the Drude
strength by non-centrosymmetry. The FS character is clarified and
low energy collective electronic excitations (plasmons) arising from the electronic
fine structure are predicted. While plasmons have been studied in conventional 
WSs,\cite{plasm1,plasm2,plasm3,plasm4} their appearance in nodal loop WSs as
well as  several other properties discussed here are distinctive when the 
special ``nodal loop semimetal'' character is encountered.

\section{Structure and Theoretical Methods}
As shown in Fig. \ref{str},
NbP has a body-centered tetragonal lattice of the space group $I4_1md$ (\#109).\cite{tian}
This space group has $4mm$ point group, is non-symmorphic, and (importantly) lacks inversion symmetry.
The structure can be pictured as columns of face-shared NbP$_6$ trigonal prisms 
oriented along the $\hat a$-axis, 
edge-shared along the $\hat{b}$-axis in one layer.
The trigonal columns are $90^\circ$ rotated in the next layer
along the $\hat{c}$-axis, giving overall tetragonal symmetry. 
In the structure with the experiment lattice parameters\cite{tian}
of $a=3.3324$ \AA, $c=11.3705$ \AA, 
both Nb and P ions lie at $4a$ positions $(0,0,z)$:
$z_{Nb}=0,$ $z_P$=0.4176; the site symmetry is $2mm$. 
Our optimized positions using the generalized gradient approximation (GGA)
with the Perdew-Burke-Ernzerhof (PBE) exchange-correlation functional\cite{pbe} 
in {\sc fplo}\cite{fplo1} coincided with the experimental values.
This structure leads to nearly identical Nb-P bond lengths of 2.53$\pm$0.01 \AA~ 
and P-Nb-P bond angles around 82$^\circ$.

Calculations with the experimental structure were performed
using the all-electron full-potential code {\sc wien2k},\cite{wien2k}
with selected results confirmed with {\sc fplo}.
All results are based on the PBE-GGA exchange-correlation functional.
One objective is to determine the combined effects of SOC and the lack of inversion
center, since delicate features around 
$E_F$ are sensitive to SOC and become important.

Calculation of optical properties including SOC is available in {\sc wien2k}.\cite{optic}
The dielectric function $\epsilon_{ij}(\omega)$, with only diagonal 
$\epsilon_{aa}=\epsilon_{bb}$ and $\epsilon_{cc}$ components due to tetragonal symmetry,
 is decomposed into the intra- and inter-band
contributions. The imaginary parts of each contribution are given by\cite{optic,valenti}
\begin{eqnarray}
 Im\epsilon_{jj}^{intra}(\omega) &\propto& \frac{\Gamma\Omega_{p,jj}^2}{\omega(\omega^2+\Gamma^2)}, \\
 Im\epsilon_{jj}^{inter}(\omega)&\propto& \sum_{c,v}
  \int d\vec{k} \frac{|\langle c_{\vec k}|p_j|v_{\vec k}\rangle |^2}{\omega^2}
  \delta(\varepsilon_{c_{\vec k}}-\varepsilon_{v_{\vec k}}-\omega), \nonumber
\label{eqn}
\end{eqnarray}
where $\vec p$ is the momentum operator, and 
$\varepsilon_{c_{\vec k}}$ and $\varepsilon_{v_{\vec k}}$ are energies of the occupied $v_{\vec k}$
and unoccupied $c_{\vec k}$ orbitals, respectively.
In the intraband term, which contains the Drude divergence for $\omega\rightarrow 0$,
$\Omega_{p,jj}$ is the Drude plasma frequency and $\Gamma$ (chosen to be 10 meV here) 
is an inverse scattering lifetime $\tau$.
The corresponding real parts are obtained by the Kramers-Kronig relation.

The transport calculations were carried out using semiclassical Bloch-Boltzmann transport theory, 
implemented in the {\it BoltzTraP} code\cite{boltz} 
that is interfaced to {\sc wien2k}. No approximations beyond the
constant scattering time approximation are made.
In all calculations performed here, the Brillouin zone was sampled with
a dense $k$-mesh up to $60\times 60\times 60$ 
to treat the semimetallic fine structure carefully.
In {\sc wien2k}, the basis size was determined by $R_{mt}K_{max}=8$,
and augmented plane-wave sphere radii of 2.5 Bohr (Nb) and 2.14 Bohr (P) were used.

\begin{figure}[tbp]
{\resizebox{8cm}{6cm}{\includegraphics{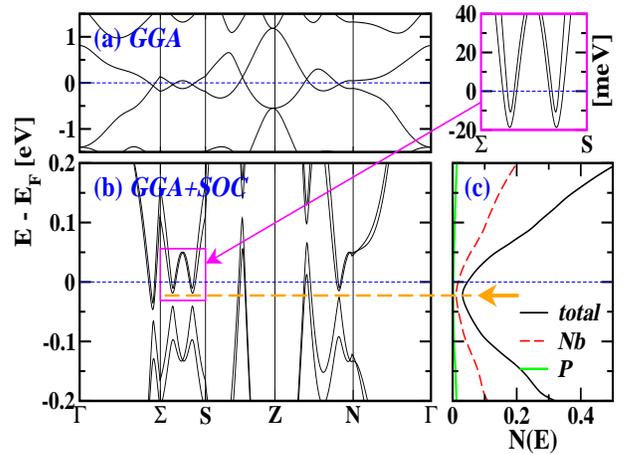}}}
\caption{(Color online) (a) GGA bands in the range of -1.5 eV to 1.5 eV 
 and (b) GGA+SOC bands in the range of -0.2 eV to 0.2 eV, with E$_F$ set to zero.
The symmetry points can be identified in Fig. \ref{fs}. The upper right panel
enlargement shows the splitting by $\sim$10 meV of bands by
SOC in this non-centrosymmetric structure. (c) Density of states (GGA+SOC)  
showing the pseudogap centered 20 meV below E$_F$.
}
\label{band}
\end{figure}

\begin{figure}[tbp]
{\resizebox{7.2cm}{14.0cm}{\includegraphics{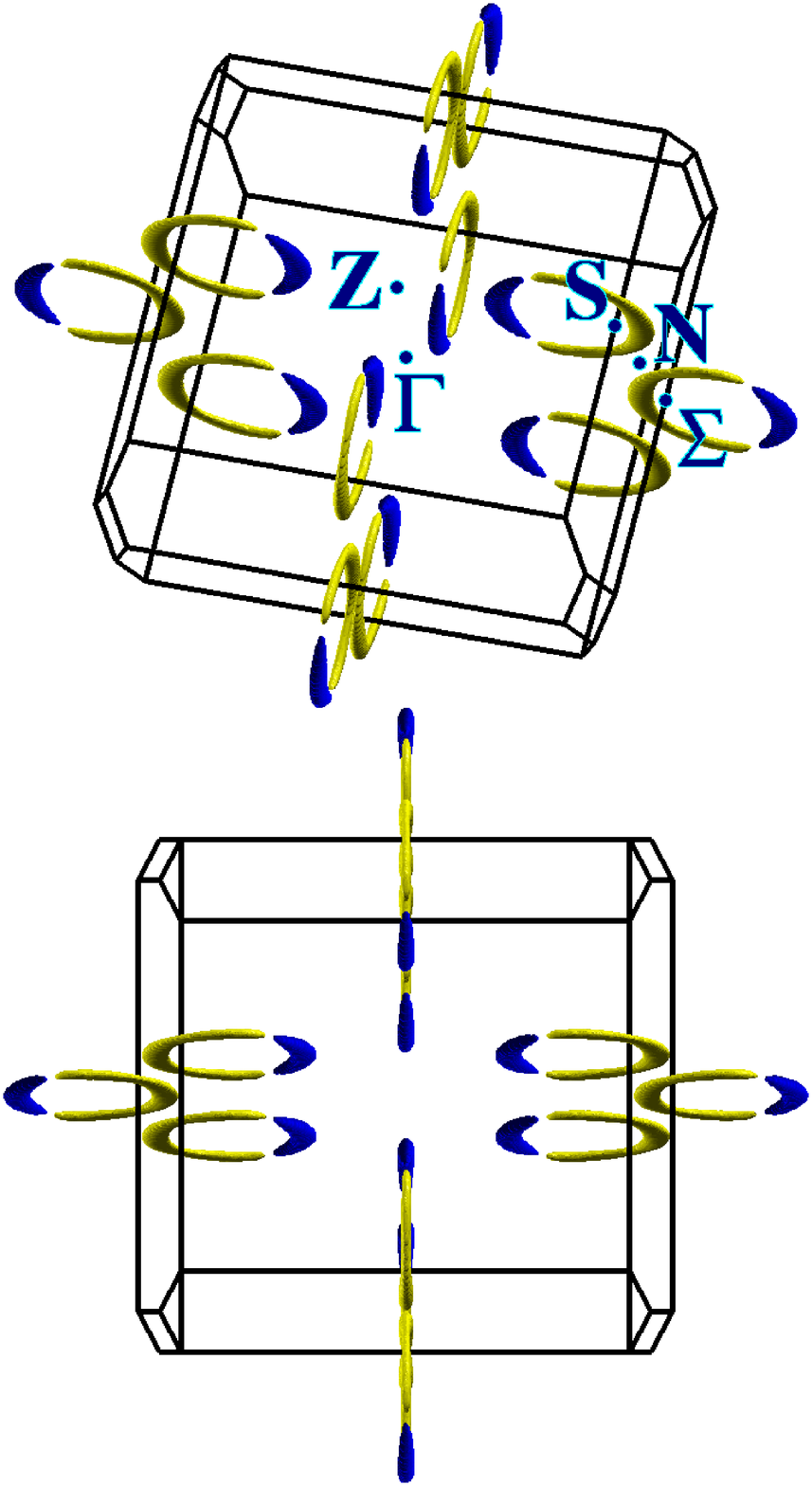}}}
\caption{(Color online) NbP Fermi surfaces (FSs) including SOC and shown
from two angles, the lower one viewed along the $k_x$ axis to reveal
the thinness of the surfaces in the mirror planes.
The extended zone representation makes the toroidal origin of the
FSs evident.
The hole (blue) boomerang FSs form bent and `sharply' tipped boomerang surfaces,
the electron (yellow) new moon surfaces are similar but extend over a larger
angle of the torus.
}
\label{fs}
\end{figure}

\section{Results}
\subsection{Electronic fine structure}
We first address the electronic structure obtained with fine $k$-point
and energy meshes.
Figure \ref{band} shows the GGA and GGA+SOC band structures near E$_F$.  
Semimetallic character with a band crossing along at least three symmetry
directions appears, with valence-conduction overlaps mostly of the order of 150 meV.
This band behavior leads to a pseudogap centered 20 meV below $E_F$ (Fig. \ref{band}),
with a very small density of states (DOS) $N(E_F)$=0.046 states/eV per f.u.
While these results are similar to the previous report,\cite{weng,she}
we examine the origins and consequences more completely.

Inclusion of SOC leads to unconventional consequences.
As usual, SOC converts many band crossings along symmetry lines into anticrossings; here
gapping bands by a few tens of meV near E$_F$. However,
the lack of inversion symmetry leads to lifting of band degeneracy over entire
symmetry planes,\cite{weng}
with splittings in the 20-200 meV range.
In spite of the modest magnitude of SOC in $4d$ Nb, 
SOC will play an important role in determining both the thermal and
spectroscopic properties of NbP.

\subsection{Fermi surfaces}
The FSs obtained from GGA+SOC are displayed in Fig. \ref{fs},
see the caption for a description.
They are similar to those presented earlier, but they are displayed in an extended
zone manner to illustrate their toroidal origin. Weng {\it et al.} have
described in detail\cite{weng} the underlying features of the Weyl nature of NbP and
isovalent partner compounds. Without SOC, mirror symmetry protected nodal loops lie precisely in
the mirror planes $k_x=0$ and $k_y$=0. The nodal loops consist of band
degeneracies $\varepsilon_{1,s}(\vec k) = \varepsilon_{2,s}(\vec k)$ ($s$
is spin degeneracy). The locus of $\vec k$ points at this intersection forms
the ring in the $\Gamma$ZN mirror plane, not at constant energy but, for these
compounds, in a small energy range including $E_F$. The FS intersects the 
nodal loop, so it must include an even number of  doubly
degenerate {\it Fermi points} where $\varepsilon_{1,s}(\vec k) = \varepsilon_{2,s}(\vec k)$ = E$_F$.
Such nodal loops have also been found in two dimensional SrVO$_3$ nanolayers.\cite{srvo3}

In this non-centrosymmetric structure, SOC splits the bands 
forming the nodal loops, leaving two 
symmetry-inequivalent pairs of Weyl points 
(12 in all) near but not on the FSs, the points being provided explicitly by 
Weng {\it et al.}\cite{weng} One lies in the 
$k_z=0$ plane, the other at a general $\vec k$ point lying near the $\Gamma$NZ
plane. The band splitting is highly anisotropic; 
each FS (Fig. \ref{fs}) consists of a flattened torus 
(annulus) pinched off into one ``boomerang'' containing holes
and one ``new moon'' containing electrons. Small doping will alter the position
of the pinching, hence changing the number of electron and hole carriers.

For a stoichiometric compound the FSs contain an equal 
number of holes and electrons,
and from calculated band fillings  we obtain 4.2$\times$10$^{-4}$ 
electron (hole) carriers per f.u.
This value corresponds to high velocity 
carriers of each sign separated on average by 
13 lattice constants in each direction.
The rms Fermi velocities are
$v_{F,aa}$=3.7, $v_{F,cc}$=1.6, in 10$^7$ cm/s,
a factor of 2.35 in anisotropy. Note that the velocities are typical of {\it metallic} 
compounds, not semimetals, due to the dispersive bands crossing E$_F$, {\it i.e.} the
Weyl character (analogous
to graphene). For the Drude plasma energy, the semimetallic value of $N(E_F)$ but 
normal metal velocities lead to $\Omega_{p,aa}$=1.0 eV, 
$\Omega_{p,cc}$=0.4 eV.

What cannot be seen in Fig. \ref{fs} is the doubling of the FSs resulting
from the interplay of SOC with the non-centrosymmetric crystalline symmetry. With the 
band degeneracy split,  Fermi surfaces 
become ``Russian doll nested'' pairs, one inside the other. This band-splitting
effect has been discussed for Pt-based superconductors\cite{Lee2005} and for
electronic properties more generally.\cite{samokhin} The band spitting aspect is
effective both in $k$-space (closely nested Fermi surfaces) 
and in energy, with consequences
discussed below.

\begin{figure}[tbp]
{\resizebox{7cm}{8cm}{\includegraphics{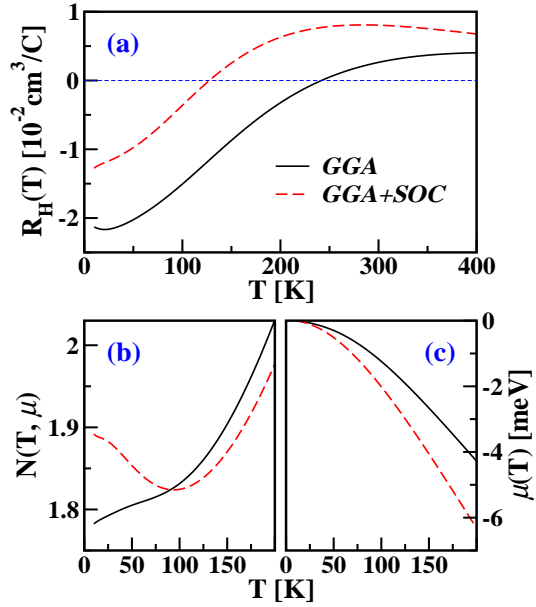}}}
\caption{(Color online) (a) Temperature (T) dependent Hall coefficients $R_H(T)$
 calculated in both GGA and GGA+SOC, which show similar T-dependence 
but distinctions in the zero crossing temperature;
T=125 K in GGA+SOC, compared to T=230 K in GGA.
 (b) T-dependent total density of states $N(T,\mu)$ 
 at chemical potential $\mu(T)$.
 $N(T, \mu)$ shows a minimum around T=100 K in GGA+SOC, while
 that of GGA increases as increasing T.
 (c) $\mu(T)$ 
 with respect to $E_F$ at T=0 K. }
\label{hall}
\end{figure}

\begin{figure}[tbp]
{\resizebox{7cm}{8cm}{\includegraphics{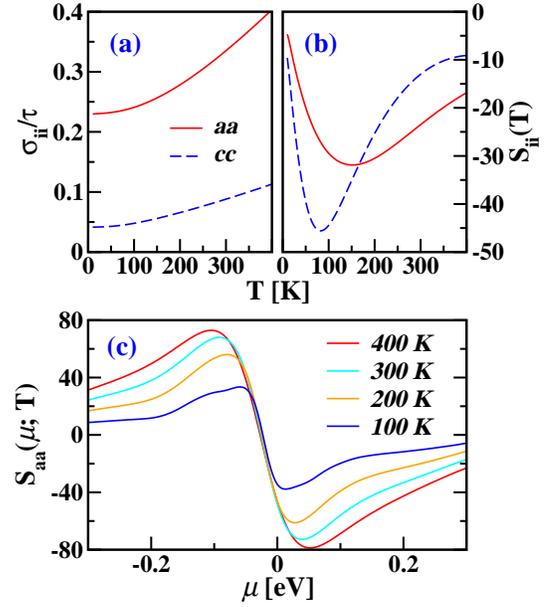}}}
\caption{(Color online)  Temperature and polarization dependent 
(a) conductivity tensor over scattering time, 
 in 10$^{20}$ ($\Omega$-m-s)$^{-1}$, and (b)
Seebeck coefficients $S_{ii}$ [$\mu$V/cm], note the minimum in the latter
in the 100-150 K
range. 
(c) The Seebeck coefficient $S_{aa}(\mu;T)$ versus band filling $\mu$
at various temperatures, within GGA+SOC. The change at very small
levels of doping is extremely rapid, reflecting the movement away from
charge compensation; $S_{cc}$ behaves in the same manner. 
The temperature variation is monotonic.
}
\label{seebeck}
\end{figure}

\subsection{Transport properties}
Transport behavior in semimetals
is very sensitive to Fermiology and to regions of high curvature,
but we have found that calculated values based on quasiclassical 
Bloch-Boltzmann theory agree very well with observed behavior.\cite{she} 
The Hall tensor components $R_H(T)$ including SOC, displayed in Fig. \ref{hall}(a),
show a change in sign at $T_c$=125 K, precisely where the experimental R$_H$
changes sign.\cite{she} This change of sign is a strong reminder that in a multiband,
compensated semimetal, R$_H$ bears no direct relation to carrier 
densities.\cite{she2} This statement is particularly true when the FSs are
strongly non-ellipsoidal and extremely anisotropic. 

The experiment-theory  agreement reflects not only the precision 
of the calculations but
also the excellent stoichiometry of the samples; the stoichiometry could not be
established precisely but is supported also by the high mobility.
The magnitude of $R_H(T)$ around 250 K is within a factor of 2 of the
experimental value. This is really excellent agreement, 
because of the compensating contributions from electron and hole surfaces, which
have regions of large curvature.  
The expression for $R_H$ in Bloch-Boltzmann theory has the interpretation of
giving it as the average of the curvature over the FSs. A strong temperature
dependence and a likely change of sign with temperature should be
anticipated in nodal loop WSs.
The experimental $R_H$ becomes at the lowest temperature
two orders of magnitude larger in magnitude than our calculated value, and
it is clear from the ``pointed'' FSs in Fig. \ref{fs} that evaluating
R$_H$ (which as mentioned is an average of the FS curvature) 
at T=0 is a substantial numerical challenge.

The chemical potential $\mu (T)$, displayed in Fig. \ref{hall}(c), decreases with 
increasing $T$ reflecting the strong differences in the electron and hole
densities of states. This variation is different in sign from another XMR compound 
WTe$_2$,\cite{ywu} reflecting different DOSs of the valence and conduction bands
of the two compounds.
The DOS at the chemical potential $N(\mu)$, shown in Fig. \ref{hall}(b), is reduced 
and achieves a minimum around $100$ K before turning around.
Figure \ref{seebeck}(a) displays the conductivity tensor components $\sigma_{ii}/\tau$, 
which incorporates the anisotropy of $\Omega_{p,jj}$, 
which is $\sim$5 times larger in-plane.

The $T$-dependent Seebeck coefficients, shown in Fig. \ref{seebeck}(b), 
have net $n$-type sign. They peak (in magnitude) at 100-150 K before again
becoming small. The dependence on chemical potential $\mu$, which can be
varied by doping or by gating, is extremely strong near stoichiometry [see Fig. 4(c)]. 
The near-perfect cancellation of electron and hole contributions (it is
perfect $S_{jj}=0$ for $\mu$$\approx$-15 meV) at
stoichiometry is destroyed, with a maximum of $|S(\mu)|$ being attained
around 0.5\% hole doping, or half of that for electron doping. The
sensitivity makes the thermopower an important gauge of the degree of
stoichiometry of samples, which otherwise can be very difficult to
determine at this low level.

\begin{figure}[tbp]
{\resizebox{7cm}{5cm}{\includegraphics{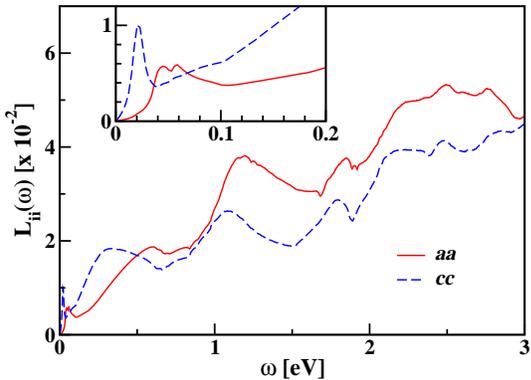}}}
\caption{(Color online) The polarization-dependent energy loss
function for NbP assuming a scattering rate $\Gamma=\hbar/\tau$
of 10 meV. Inset: the $\hat c$-axis plasmon at 20 meV bifurcates into a
pair of plasmons at 40 meV and 60 meV for in-plane polarization. 
Above 0.2 eV the loss
function is not greatly anisotropic and is insensitive to $\Gamma$. 
}
\label{optic}
\end{figure}

\subsection{Optical properties}
The band splitting induced by SOC and the lack of inversion symmetry have
qualitative consequences for the optical properties. In {\it conventional}
WSs where Fermi surfaces are simple and model band structures work well,
the dielectric response and plasmon signatures of chirality have been
well studied.\cite{plasm1,plasm2,plasm3,plasm4} Nodal loop WSs bring
new processes to the low-energy, small-$q$ dielectric response.
They are compensated semimetals thus providing itinerant electron and
hole gases. They are qualitatively different from other semimetals
(including conventional WSs) by having highly non-parabolic and
extended Fermi surfaces, and of course by having very closely nested
Fermi surfaces. New dielectric phenomena must emerge.
 
Without either SOC or inversion symmetry breaking, bands are doubly
degenerate and contribute accordingly to the intraband Drude response.
In nodal loop WSs half of this weight is transferred to finite-$q$,
finite-$\omega$ response. The changes will be appreciable only at small
energies, and they should involve new response at very small $q$ wavevectors that connect
the finely nested Fermi surfaces, which are on the order of 10$^{-2}$$\pi/a$.
The Drude response will be relatively unchanged 
apart from the factor of 2 reduction in weight, but it may be smeared into
the new small-$\omega$, small-$q$ structures.

The energy loss function,
shown in Fig. \ref{optic} with scattering rate $\Gamma$=10 meV, 
displays low energy plasmons when SOC is included,
at 20 meV for $\hat c$-axis polarization and double in-plane plasmon peaks
at 40 meV and 60 meV. The strength of these peaks, and to some degree their
positions, are dependent on the presumed scattering rate $\Gamma$. 
The anisotropy we calculate at low energy, 
based on the strong anisotropy of the Fermi
surfaces, may be consistent with the angle-dependence
of the MR measurements.\cite{she}
Given the reported high mobility, the plasmon peaks may be much sharper at low
temperature than calculated here. The peaks can be
expected also to have temperature dependence, because the varying $\mu(T)$
induces changes and $\Gamma$ will be T-dependent as well. Above 200 meV, we
do not expect significant T-dependence of the spectral behavior.
Far-infrared spectroscopy may reveal unusual behavior as temperature, field,
polarization, and carrier density are varied and tuned.

\section{Discussion and Summary}
Our study has illuminated the origins of the unusual and delicate Fermi surfaces of NbP.
Without consideration of spin-orbit coupling, the electronic structure includes
nodal loops\cite{weng} lying in the mirror planes that lead to point Fermi surfaces
as well as small electron and hole pockets. In this non-centrosymmetric
structure, spin-orbit coupling removes the
spin degeneracy, resulting in Russian doll nested pairs of Fermi surfaces with 
a pinched-off annulus topology, with electron surfaces converting to hole
surfaces across the pinch. The resulting electronic fine structure accounts
well for the change in sign of the Hall coefficient at 100 K, which is also
the temperature range in which the magnetoresistance begins to grow in size.
These features identify 10 meV as a relevant energy scale.

The unusual Fermi surface also accounts for the lack of universal scaling of
XMR in NbP and TaP; spin-orbit coupling is much larger in TaP and the Fermi
surfaces are sensitive to this.
Finally, the fine structure in NbP results in the 
appearance of polarization dependent low 
energy plasmons in the 20-60 meV range. This study substantially broadens
the understanding of how WS behavior impacts the physical
properties of NbP, which should also apply in similar form
to NbAs, TaP, and TaAs.

\begin{acknowledgments}
 We acknowledge J.-G. Hong for useful communications on magnetoresistance
 and A. S. Botana for useful discussion on calculations in BoltzTraP.
 This research was supported by National Research Foundation of Korea
 Grant No. NRF-2013R1A1A2A10008946 (K.H.A and K.W.L),
 by U.S. National Science Foundation Grant DMR-1207622-0 (K.W.L.)
 and by U.S. Department of Energy Grant DE-FG02-04ER46111 (W.E.P.).
\end{acknowledgments}

\end{document}